\documentclass[prb,twocolumn,showpacs,superscriptaddress]{revtex4}
\usepackage{graphicx,dcolumn,amsmath}

\begin{document}

\title{\bf
Electron confinement and optical enhancement in Si/SiO$_2$ superlattices
}

% Authors and affiliations
\author{\firstname{Pierre} \surname{Carrier}}
\affiliation{D\'{e}partement de Physique et Groupe de Recherche en
Physique \\et Technologie des Couches Minces (GCM), Universit\'{e} de
Montr\'{e}al, \\ Case Postale 6128, Succursale~Centre-Ville, Montr\'{e}al,
Qu\'{e}bec, Canada H3C 3J7}

\author{\firstname{Laurent J.} \surname{Lewis}}
\email[To whom correspondence should be addressed; \\
e-mail: ]{laurent.lewis@umontreal.ca}
\affiliation{D\'{e}partement de Physique et Groupe de Recherche en
Physique \\et Technologie des Couches Minces (GCM), Universit\'{e} de
Montr\'{e}al, \\ Case Postale 6128, Succursale~Centre-Ville, Montr\'{e}al,
Qu\'{e}bec, Canada H3C 3J7}

\author{\firstname{M. W. Chandre} \surname{Dharma-wardana}}
%\email[E-mail: ]{chandre@cm1.phy.nrc.ca}
\affiliation{Institute for Microstructural Sciences, National Research
Council, Ottawa, Canada K1A 0R6}

\date{\today}

\begin{abstract}
We have performed first-principles calculations of Si/SiO$_2$ superlattices
in order to examine their electronic states, confinement and optical
transitions, using linearized-augmented-plane-wave techniques and
density-functional theory. Two atomic models having fairly simple interface
structure are considered; they differ in the way dangling bonds at
interfaces are satisfied. We compare our first-principles band structures
with those from tight-binding calculations. The real and imaginary parts of
the dielectric function are calculated at the Fermi-Golden-rule level and
used to estimate the absorption coefficients. Confinement is confirmed by the
dispersionless character of the electronic band structures in the growth
direction. Optical enhancement is shown to exist by comparing the direct and
indirect transitions in the band structures with the related transitions in
bulk-Si. The role played by the interface on the optical properties is
assessed by comparing the absorption coefficients from the two models.
\end{abstract}


\maketitle

\section{Introduction}\label{Intro}

Informatics and telecommunications are the driving forces behind the
development of new photonic devices.\cite{lampro} Such devices can be
assembled from a wide variety of materials; among these, silicon is probably
the most prevalent, although it suffers from having an indirect band gap.
Direct band-gap materials such as GaAs, InP, and many other III-V and II-VI
semiconductors, are efficient electron-photon energy converters. However they
are costly for large-scale applications, and would be less interesting if
full Si-based photonics were possible. These considerations have lead to
considerable efforts towards the development of new types of Si-based
materials having direct energy gap in the visible (viz.\ 1.5 to 3
eV),\cite{Zhang} suitable for photo-diodes and lasers, as well as solar-cell
applications.

The simultaneous discovery by Canham\cite{Canham} and Lehmann and
G\"{o}sele\cite{LehmannGosele} of intense luminescence in porous silicon
($\pi$-Si) has opened up new horizons for Si-based materials. $\pi$-Si can be
regarded as consisting of long and thin nanowires,\cite{ChuangCollinsSmith}
forcing the electronic states to be confined within the finite dimension of
the nanostructures. The shift of the luminescence spectrum towards the blue
with decreasing nanowire size\cite{Canham} suggests that the confinement of
the electronic states might be responsible for the visible luminescence.
Unfortunately, most forms of $\pi$-Si are unstable, lasting only a few hours
before a strong decrease of the luminescence efficiency is observed. This
instability has sometimes been attributed to the volatile Si--H bonds at the
surface.\cite{LehmannGosele} Annealing under nitrogen and oxygen has been
suggested as a means of stabilizing the structures; however, the annealed
material has weaker luminescence than the parent $\pi$-Si. Layered
polysilanes (Si$_{6n}$H$_6$) also exhibit optical properties similar to
$\pi$-Si. Uehara et al\cite{Uehara} have performed first-principles
calculations of layered polysilanes structures and showed that porous Si may
be akin to thin Si$_6$H$_6$ sheets.

As a perhaps more promising avenue, confinement can also be achieved in
superlattices (SLs) --- or quantum wells --- as well as quantum wires and
quantum dots. Lu et al have reported enhanced luminescence in the visible
part of the spectrum of Si/SiO$_2$ SLs fabricated by molecular beam epitaxy
(MBE),\cite{LuLockwoodBaribeau} and a blue shift was observed when the
thickness of the Si layers was reduced from 6 to 2 nm. In contrast to
$\pi$-Si, Si/SiO$_2$ SLs present the advantage of being stable, with no
significant decrease of the luminescence observed with time.

Motivated by these promising experimental observations, and to provide
insight into the microscopic physics associated with the luminescence
efficiency of SL-based devices, we undertook a detailed, first-principles
investigation of the electronic and optical properties of Si/SiO$_2$ SLs.
Silicon and SiO$_2$ are both already standard components of MOSFETs and other
devices. SiO$_2$ has an energy gap of $\sim$9 eV; Si has an indirect gap of
$\sim$1.1 eV. Thus, the electrons are confined by the SiO$_2$ barriers within
the silicon quantum wells. Luminescence, however, is not determined solely by
confinement since the nature of the energy gap is also important. The
possibility of creating a direct energy gap in a SL can be discussed in
simplified terms within the concept of Brillouin zone
folding:\cite{gnutzmann} The periodicity of a SL, $d$, being larger than that
of the constituent lattices, $a_0$, the Brillouin zone gets folded to
dimension $\pi/d$ instead of $\pi/a_0$, which may bring the indirect minimum
of the conduction band to the $\Gamma$-point and thus produce an effectively
direct energy-gap material. ``Quasi-direct'' energy gaps were so obtained in
Si/Ge superlattices;\cite{pearsall} however, there was little enhancement in
the optical properties.

In this work, we examine the luminescence properties of Si/SiO$_2$ SLs using
model structures and first-principles methodology. Transmission electron
microscopy reveals that in Si/SiO$_2$ SLs, both Si and SiO$_2$ layers are
amorphous.\cite{LuLockwoodBaribeau} Amorphous structures are however not
easily amenable to quantum calculations. Here, we assume crystalline phases
for both the Si and SiO$_2$ slabs. This is further motivated by the need to
study simple ideal systems as benchmarks, and in particular to determine how
the difference in the interface affects the optical properties. We however
note that crystalline-Si based SLs have very recently been
fabricated.\cite{Lu_private}

\begin{figure}[t]
\includegraphics*[width=9cm]{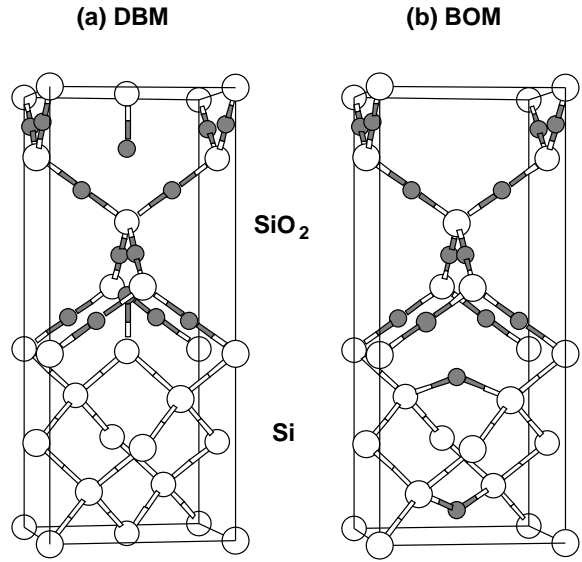}
\caption{
The two model structures considered in the present study: (a) the DBM,
which contains 23 atoms (13 Si and 10 O), and (b) the BOM, which contains 21
atoms (11 Si and 10 O atoms). Both unit cells are rectangular, of size
$5.43\times5.43\times13.11$ \AA$^3$. The two models differ only by
their Si-SiO$_2$ interface.
}
\label{modelFIG}
\end{figure}

Two simple model structures for Si/SiO$_2$ SLs are considered: (i) The
double-bonded model (DBM) of Herman and Batra,\cite{HermanBatra} shown in
Fig.~\ref{modelFIG}(a); here the Si and SiO$_2$ layers are arranged so as to
minimize the lattice mismatch and have relatively high symmetry, and the
dangling bonds at the interfaces are saturated by the addition of a single
double-bonded oxygen atom. (ii) A bridge-oxygen model (BOM), shown in
Fig.~\ref{modelFIG}(b), where the interface dangling bonds are saturated by
adding an oxygen atom to bridge two Si bonds. Full details of the models are
given in Section~\ref{StructSec}.
%FIGURE_HERE

The electronic properties of both the DBM and the BOM have been studied
already within an empirical tight-binding framework by Tit and
Dharma-wardana,\cite{TitDharma} who observed the bands along the
high-symmetry lines of the BZ parallel to the $k_z$ axis to be essentially
dispersionless, in agreement with the present calculations. However, the band
gaps were found to be direct in both models, in contradiction with our
results. The DBM has been studied from first principles by Punkkinen et
al.\cite{Punkkinen} using a full-potential linear-muffin-tin orbital
approach. However, no other interface-structure was examined and thus the
effect of changing the interface could not be assessed; further, the optical
properties were not calculated. Kageshima and Shiraishi have also used
first-principles methods to examine different models for the Si--SiO$_2$
interface in oxyde-covered Si slabs,\cite{Kageshima} but the SL geometry was
not examined. Their calculations indicate that the interface plays an
important role in the light-emitting properties of the system and that, in
particular, interfacial Si-OH bonds are possibly related to the observed
luminescence. Here we present results for the electronic structure, the
dielectric function and the absorption coefficient of both the DBM and the
BOM, and focus on a comparison between the two models.

\section{Computational details}\label{CompDet}

\subsection{Model structures}\label{StructSec}

SLs can be fabricated by MBE;\cite{HermanSitter} a major advantage of this
technique, as compared to others, is that it gives sharper interfaces. In the
case of Si/SiO$_2$ SLs, Lu et al.\cite{LuLockwoodBaribeau} report an
interface thickness of about 0.5 nm, while the thickness of the SiO$_2$ and
Si layers vary between 2 and 6 nm. Here we assume that the interfaces are
perfectly sharp.

The model Si/SiO$_2$ SL structures are constructed by alternating layers of
crystalline (diamond) silicon and crystalline SiO$_2$ in the ideal
$\beta$-cristobalite structure, viz. with a Si--O--Si bond angle of
180$^{\circ}$ (space group Fd$\bar{3}$m). The (experimental) lattice
constants are 5.43 and 7.16 \AA, respectively, with therefore a lattice
mismatch of 32\%. The mismatch can be reduced to less than 7\% by rotating
the Si unit cell by an angle of $\pi/4$ so that the diagonal of the
Si-diamond structure fits, approximately, the cubic edge of the
$\beta$-cristobalite unit cell. The superposition of the Si and SiO$_2$
layers gives rise to dangling bonds on some of the interface Si atoms. These
can be satisfied in different ways. In the DBM, this is done by adding a
single O atom onto one of the Si atoms. The resulting unit cell has 23 atoms
[cf.\ Fig.\ref{modelFIG}(a)] and dimensions 5.43 \AA\ in the $x-y$ plane and
5.43~$\times$~(1~+~$\sqrt{2}$)~=~13.11~\AA\ in the $z$ (growth) direction. In
this model, the Si--Si and Si--O distances, 2.35 and 1.66 \AA, respectively,
are realistic, but the double-bonded oxygen Si=O is not found in naturally
occuring silicates. In the BOM, Fig.~\ref{modelFIG}(b), an oxygen atom
saturates two dangling bonds provided by two different Si atoms; this model
contains 21 atoms. Here the Si--O--Si angle is the usual 144$^\circ$ and the
length of the Si--O bonds at the interface is 2.02 \AA, i.e., somewhat longer
than found in nature. No energy relaxation of the models have been carried
out. However, as will be discussed in section~\ref{gapSec}, we found gap
states to be present in the BOM; in order to understand this, we investigated
two other variants of the BOM, having Si--O--Si angles of 109 and
158$^\circ$, respectively.

\begin{figure}[t]
\includegraphics*[width=7cm]{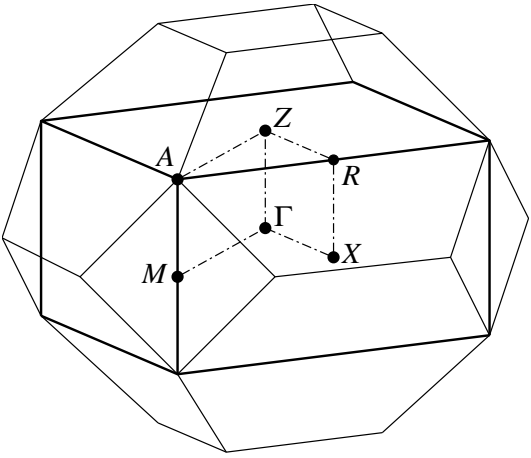}
\caption{
Brillouin zone for the SL compared to that for the diamond structure. The
principal axes of symmetry used in the electronic structures calculations are
also shown.
}
\label{bz}
\end{figure}

The Brillouin zone for the SL, along with that for c-Si, is depicted in
Figure~\ref{bz}. The principal symmetry axes are indicated: the $Z-\Gamma$,
$X-R$ and $M-A$ directions correspond to possible growth directions where
confinement is expected to occur, as discussed below.
%FIGURE_HERE

\subsection{First-principles calculations}\label{NumericSec}

The electronic-structure calculations were carried out with the {\sf WIEN97}
code,\cite{WIENref} which uses the all-electron, full-potential,
linearized-augmented-plane-wave (LAPW) method,\cite{Singh} within the
framework of density-functional theory\cite{HohenbergKohn,KohnSham} in the
local-density approximation (LDA).\cite{Payne} Thus, the core states in the
(spherical) atomic region are described using an atomic basis, while the
delocalized states in the interstitial region are expanded in plane waves
which are correctly formulated to account for the core region. The Kohn-Sham
(KS) basis inside the atomic spheres is expressed as an angular-momentum
expansion:
   $$
   \psi(\vec{k}_n,\vec{r}) =
       \sum_{l,m} \left[ A_{lm}^n u_{l}(r,E_l)+B_{lm}^n
               \left.\frac{\partial u_{l}}{\partial E}\right|_{E=E_l}
                        \right] Y_{lm}(\hat{\boldmath r}).
   $$
Here the energy $E_l$ is calculated and fixed at the first cycle of the
self-consistent-field calculation. The KS equations are solved for a grid of
$\vec {k}$ points as discussed below.

The integration of the radial secular equations was performed on a radial
mesh containing 581 points. The radius of the spheres for silicon and oxygen
atoms were set at 0.87 and 0.79 \AA, respectively, and an energy cutoff of
$-7.1$ Ry was used for separating core from valence electrons. This generated
11~149 plane waves and the matrix size for the eigenvalue problem contained
5581 elements. These two parameters correspond to an angular momentum cutoff
of 7.5 (see Singh\cite{Singh} for further details), a value which ensures
proper energy convergence of the solution.

The DBM has nine non-equivalent atoms while the BOM has eight. The
irreducible wedge of the Brillouin zone is obtained after four symmetry
operations, which reduces considerably the computational load. The supercell
in the $z$ direction being 3 times longer than in the $x$ or $y$ directions,
only 181 $\vec {k}$ points were necessary to achieve convergence of the
electronic energies. However, in order to ensure convergence of the wave
functions needed to set up the optical matrix, the density of $\vec {k}$
points had to be doubled. Integration of the Brillouin zone was performed
using the tetrahedron method and Broyden density mixing. The self-consistent
calculations were performed in parallel mode (using four processors) on a
Silicon Graphics Origin 2000 computer system. About 10 cycles were necessary
to get convergence of the energy to about 10$^{-4}$ Ry.

\section{Results and discussion}\label{Results}

\subsection{Band structures}\label{BandSec}

The electronic band structures along the high symmetry axes of the tetragonal
Brillouin zone (cf.\ Fig.\ref{bz}) for the two models are plotted in
Fig.~\ref{DBMBOMband}, together with the densities of states (DOS). For the
BOM, unless otherwise noted, the interfacial Si-O-Si angle is the normal
144$^{\circ}$. The confinement in the $z$-direction can already be inferred
from the essentially dispersionless character of the bands along the three
symmetry axes which are parallel to the axis of the SL, namely $X-R$,
$Z-\Gamma$ and $M-A$; this was also observed by Punkkinen et
al.\cite{Punkkinen} Thus, optical transitions between bands lying in the
growth direction are favored by the confinement.
%FIGURE_HERE

The DOS of the two models differ strongly in the region of the gap, with an
isolated conduction band clearly visible in the gap of the BOM which is
absent in the DBM. The only difference between the two models lies in the way
that the interfacial dangling bonds are fixed; the gap state in the BOM is
thus connected to the bridge-bonded Si atom. The results, evidently, are
sensitive to the way that the interface is formed, either in experiment, or
in a theoretical model. This question will be discussed in more detail in
section~\ref{gapSec}.

The effect of the confinement in both models is clearly visible from the DOS.
If we ignore the gap state in the BOM for the moment, the DOS at the Fermi
level has a step-function-like threshold, as opposed to bulk-silicon which is
known\cite{Cardona} to have a threshold depending on the photon energy
$\hbar\omega$ as $(\hbar\omega-E_g)^{1/2}$, where $E_g$ is the gap energy.
Higher transition probabilities are thus expected in Si/SiO$_2$ SLs.

Figure~\ref{fermiband} zooms on the band structure of the two models near the
Fermi level. The numerical values of the various possible transition energies
are given in Table~\ref{gaps}. For the DBM, the valence-band maximum (VBM) is
at the $\Gamma$ point but the conduction band minimum (CBM) is in between the
$\Gamma$ and $X$ points (labeled ${\Gamma X/2}$ in Table~\ref{gaps}), giving
an indirect gap of 0.81 eV. For the BOM, the VBM is at $X$ while the CBM is
at $Z$, giving an almost null (0.01 eV) indirect band gap. However, it should
be remembered that the energy gaps are significantly underestimated in the
LDA and other implementations of the DFT. Thus the bulk-Si LDA energy gap is 0.49
eV, about 0.6 eV below the true value of 1.1 eV. Better (but non-rigorous)
estimates for the DBM and the BOM energy gaps would thus be 1.4 and 0.6 eV,
respectively. The unusually low value for the energy gap of the BOM is an
indication that the lowest conduction band is really a gap state. If this
state [``CB1'' in Fig.\ref{fermiband}(b)] is assumed to be an unphysical
artifact of the model, and temporarily ignored, the indirect gap of 1.49 eV
gets corrected to $\sim$2.1 eV. Thus, the energy gap would be in the near
infrared or in the red, consistent with
experiment.\cite{LuLockwoodBaribeau,Mulloni,Novikov}
%FIGURE_HERE

The dispersionless character of the bands in the growth direction provides
evidence of the confinement effect, as noted above. However, confinement
alone is not sufficient to explain the enhancement of the optical properties;
comparison of the direct and indirect transitions for the two models can
clarify this question. For the DBM, the direct transitions at $\Gamma$ and
$\Gamma X/2$ have energies of 0.98 eV and 0.84 eV, respectively, while the
$\Gamma-\Gamma X/2$ indirect transition (i.e., the band gap) costs 0.81 eV.
Thus the {\em smallest direct--indirect gap} (SDIG) in the LDA for the DBM is
a mere 0.03 eV ($0.84-0.81$ eV). For the BOM (if the ``CB1'' gap state is
ignored), the direct transitions at $X$ and $Z$ have energies of 2.19 and
1.87 eV, respectively, compared to 1.49 eV for the $Z$--$X$ indirect
transition, for a SDIG of only 0.35 eV. These numbers can be compared to the
corresponding ones for bulk silicon (in the LDA): the direct transition at
the $\Gamma$ point has energy 2.52 eV while the direct transition at the CBM
(near $3/4\Gamma X$) costs 3.20 eV. The indirect band gap is 0.49 eV, so that
the SDIG in this case is $\sim$2.0 eV. This is {\em much larger} than that
for the DBM and the BOM. The latter, as a consequence, has much better
optical properties than bulk Si. It is also clear that the DBM will have
better optical properties than the BOM (see section~\ref{OpticSec}) since the
SDIG of the DBM is smaller than that of the BOM.

\subsection{Comparison with tight-binding calculations}\label{TBsec}

It is instructive to compare the present results to those obtained for the
same model systems by Tit and Dharma-wardana\cite{TitDharmaDBMBOM} within a
tight-binding (TB) framework. In a TB description,\cite{Papaconstantopoulos}
the wavefunctions are expressed as linear combinations of atomic orbitals and
the exact many-body Hamiltonian is replaced by a parametrized Hamiltonian
matrix whose eigenvalues and eigenvectors yield the energies and the
wavefunctions of the corresponding electron levels. The size of the numerical
problem is determined by the number of atomic orbitals chosen to describe the
valence electrons, and about 4 or 5 orbitals per atom are typically used.
This allows the study of much larger systems than is possible in
first-principles calculations. It is therefore important to assess the
validity of TB models in order to re-calibrate and improve the TB
parametrization. Indeed, the parameters are typically fitted to experimental
data or first-principles calculations for the {\em pure} crystalline phases.
For mixed systems, e.g., a SL, the parameters are assumed to be transferable
except for simple adjustments of energy reference or scaling to bond lengths,
etc.

An important parameter entering TB calculations is the valence-band offset
(VBO), which is the difference in energy between the VBM of the two materials
constituting the SL.\cite{Williams} In their calculations, Tit and
Dharma-wardana assumed a VBO of 3.75 eV, based on the experimental offset
between Si and amorphous SiO$_2$. However, this value need not be appropriate
to the idealized DBM and BOM models.
%FIGURE_HERE

\begin{figure}[t]
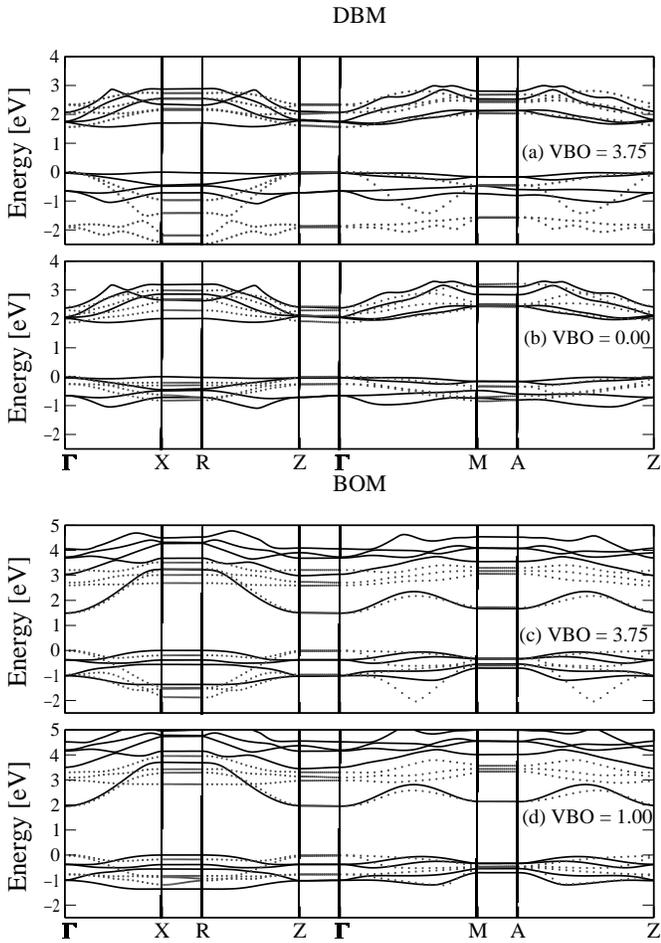

\includegraphics*[width=8.7cm]{Figs/VBO_DBM.eps}
\hspace{1cm}
\includegraphics*[width=8.7cm]{Figs/VBO_BOM144.eps}
\caption{
Comparison between TB (dotted lines) and first-principles (full lines) band
structures for the two models, for two different values of the VBO: (a) 3.75
eV and (b) 0.0 eV for the DBM; (c) 3.75 eV and (d) 1.0 eV for the BOM. The
LDA energy gaps were ``manually'' set to the gap from TB calculations in
order to facilitate the comparison.
}
\label{tbvswien}
\end{figure}

In Fig.~\ref{tbvswien}(a) we compare the LDA band structure for the DBM with
the TB results of Tit and Dharma-wardana,\cite{TitDharmaDBMBOM} who used a
VBO value of 3.75 eV; Fig.~\ref{tbvswien}(c) is the corresponding plot for
the BOM. The agreement with this particular value of the VBO is, at best,
qualitative; this may indicate that the choice of VBO needs revision. In
order to assess this, we have done new TB calculations using different VBO
values. By adjusting as closely as possible the valence band structures to
the TB ones, better values of the VBO were found to be 0.0 eV for the DBM and
1.0 eV for the BOM; the results are indicated in Fig.~\ref{tbvswien}(b) and
(d), where the band gaps from LDA are adjusted to the TB values to help
comparison. For the BOM, the gap state ``CB1'' from LDA coincides well with
the first conduction band obtained from TB, except in the $X$--$Z$ region of
the BZ. The first-principles ``CB2'', in contrast, is somewhat higher than
found from the TB calculations. Thus, while not perfect, the agreement with
the first-principles calculations improves significantly with these new VBO
values.

\subsection{Gap states in the BOM}\label{gapSec}

In view of the presence of a gap state in the BOM --- labeled ``CB1'' in
Fig.~\ref{fermiband} --- and in order to assess the role of the interface on
the electronic properties, other possibilities for the interfacial Si--O--Si
angle were considered: (i) 109$^{\circ}$, corresponding to positioning the
oxygen atom on a normal silicon site of the Si lattice; here, $d_{\rm Si-O}=
2.35$ \AA; (ii) 144$^{\circ}$ which corresponds to the experimental value of
the Si--O--Si angle, used in the calculations discussed above, yielding
$d_{\rm Si-O} = 2.02$ \AA; and finally (iii) 158$^{\circ}$, a value obtained
by relaxing the oxygen atom position, and which corresponds to a (local)
energy minimum; in this case, $d_{\rm Si-O} = 1.96$ \AA. The ``normal'' Si--O
distance in silica is 1.61 \AA, which cannot be accomodated by the
crystalline silicon lattice in the BOM.

\begin{figure}[t]
\includegraphics*[width=8.7cm]{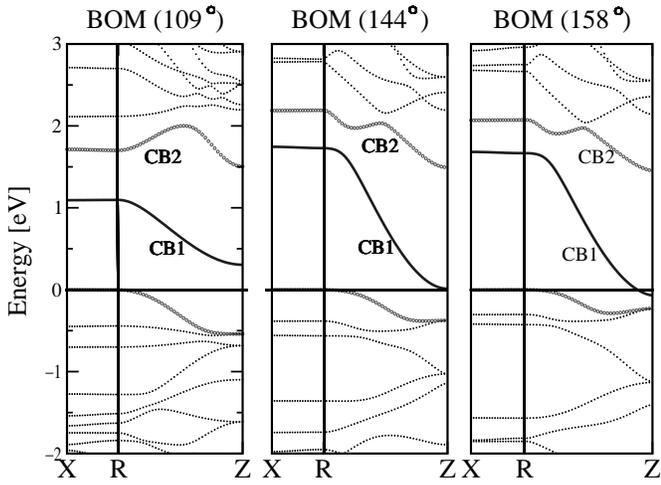}
\caption{
Band structure of the BOM having three different interfacial Si--O--Si angles
(as indicated).
}
\label{gapstatefig}
\end{figure}

Figure~\ref{gapstatefig} shows the band structures for the three cases; here
we consider only the $X-R-Z$ direction, wherein lies the energy gap.
Evidently, the precise value of the interfacial Si--O--Si bond angle, and
corresponding Si--O bondlength, have a sizeable effect on the band structure.
In fact, the 158$^{\circ}$--BOM is, within the LDA, a metal if CB1 is not
assumed to be a gap state, as discussed earlier. Likewise, the
144$^{\circ}$--BOM is nearly metallic. These results indicate that the
acceptable range of interfacial Si--O--Si angles, and the consequent longer
than ``normal'' Si--O distances, are probably essential issues in explaining
the luminescence in Si/SiO$_2$ SLs.
%FIGURE_HERE

\subsection{Optical properties}\label{OpticSec}

The Kohn-Sham calculations provide matrix elements and joint densities of
states necessary for the calculation of the complex dielectric function
$\vec{\epsilon} = \vec{\epsilon}_r +i\vec{\epsilon}_i$. The absorption
coefficient $\alpha$, which can then be deduced from $\vec{\epsilon}$ as a
function of the photon energy (see below), provides a detailed picture of one
aspect of the optical properties of the material. In principle, the
luminescence requires a knowledge of the excited states of the system, with
electrons occupying the conduction band. Such states are not available from
DFT calculations. Further, the electrons in the conduction band are
associated with holes in the valence band. For low carrier concentrations,
the screening is weak and hence exciton formation occurs.\cite{knox} This
many-body effect is also required for a complete description of luminescence.
Such calculations require, e.g., the solution of the Bethe-Salpeter equation
for the electron-hole pair (see Chang et al., Ref.\ \onlinecite{Chang}) and
are still prohibitive even for our model structures. Excitonic effects will
thus be neglected here.

As already noted, the band gaps are underestimated by LDA-DFT and this has to
be corrected if the absorption thresholds are to be realistic. In practice,
this can be done to a reasonable approximation by rigidly shifting all the
conduction bands to the appropriate energy. However, we remain within the DFT
framework and calculate the absorption in the Fermi-golden-rule
approximation. Also, we assume that the emission spectrum is similar to the
absorption spectrum and neglect excitonic effects. In spite of these
approximations, the main mechanisms of luminescence enhancement --- the
effects arising from the joint density of states and the matrix elements ---
would be correctly captured.

Using the program written by Abt et al.\cite{AbtDraxlKnoll} as part of the
{\sf WIEN97} package,\cite{WIENref} the imaginary part of the dielectric
function has been determined; it is given by:
   \begin{eqnarray*}
   \epsilon_i^{\alpha}(\omega) = & \left(\frac{4\pi e^2}{m^2\omega^2}\right)
   \sum_{v,c} \int  \frac{2d\vec{k}^3}{(2\pi)^3} |\!\!<\!\!c\vec{k}
   | {\cal H}^{\alpha} | v\vec{k}\!\!>\!\!|^2 \\ \nonumber
   & \!\!<\!\! f_{v\vec{k}}(1\!\!-\!\!f_{c\vec{k}}) \delta(E_{c\vec{k}} - E_{v\vec{k}} -
   \hbar \omega), \\ \nonumber
   \end{eqnarray*}
where $f_{v\vec{k}}$ is the Fermi distribution and ${\cal H}^{\alpha}$ is the
$\alpha$-component of the electron-radiation interaction Hamiltonian in the
Coulomb gauge; it corresponds to the probability per unit volume for a
transition of an electron in the valence band state $|v\vec{k}\!\!>$ to the
conduction band state $|c\vec{k}\!\!>$ to occur. From Kramers-Kronig
relations one then deduces the real part $\epsilon_r$.\cite{Cardona} The
dielectric function $\vec{\epsilon}$ is defined as the square of the complex
refractive index $\vec{n} = n_r + i n_i$, so that
   $$
   \alpha(E)= 4\pi \frac{E}{hc} n_i =
   4\pi \frac{E}{hc} \left[\frac{(\epsilon_r^2 -\epsilon_i^r)^{1/2} -
   \epsilon_r}{2}\right]^{1/2},
   $$
with $c$ the speed of light in vacuum, $h$ Planck's constant, and $E$ the
photon energy.

\begin{figure}[t]
\includegraphics*[width=8.7cm]{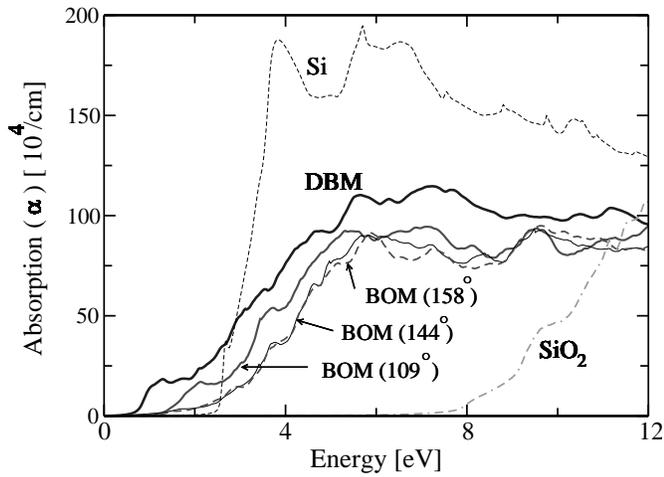}
\caption{
Absorption curves for the two models compared to bulk Si and ideal
$\beta$-cristobalite SiO$_2$. The three BOM angles correspond to the
interfacial Si--O--Si angle in the model. The BOM curves for 144$^{\circ}$
and 158$^{\circ}$ overlap almost exactly; the full line corresponds to an
angle of 144$^{\circ}$ while the dotted line corresponds to an angle of
158$^{\circ}$.
}
\label{tousabs}
\end{figure}

Figure~\ref{tousabs} shows the $z$-component of the absorption coefficients
for the DBM and for the three different variations of the BOM. We also give,
for comparison, the corresponding curves for silicon and for the ideal
$\beta$-cristobalite SiO$_2$ structures. The energy gaps are those directly
from the LDA; correcting the gaps would only affect the position of the
onset, but not the general aspect. The absorption coefficient for the two
models are similar. This similarity is, to a large extent, a ``zone folding''
effect, as the two model structures have identical $z$-dimensions (the growth
axis). However, closer inspection reveals that the DBM has better absorption
properties than the three BOMs: the onset of absorption is sharper, and
occurs earlier in the DBM than in all the BOM. This is consistent with the
band structure analysis above that showed the SDIG to be smaller in the DBM
than in the BOM ($\sim$ 0.03 vs $\sim$0.35 eV), implying a larger transition
probability for the former than the latter.

In addition, the calculated SDIG for the BOM with an interfacial Si--O--Si
angle of 109$^{\circ}$ is 0.18 eV --- the band gap (or the indirect
transition) being 1.53 eV while the direct transition between $X$ and $Z$ are
2.06 eV and 1.71 eV respectively --- in between the SDIG for the DBM and the
BOM with 144$^{\circ}$. This is in agreement with the absorption curves of
Fig.~\ref{tousabs}, where the absorption onset is higher for the DBM, lower
for the BOM with 109$^{\circ}$ and then even lower for the BOM with
144$^{\circ}$ or 158$^{\circ}$. The last two give similar absorption curves
as can be seen from Fig.~\ref{tousabs}.
%FIGURE_HERE

Thus, the four models (the DBM and the BOM with bond angles 109$^{\circ}$,
144$^{\circ}$ and 158$^{\circ}$), which differ only by their interface
definition, give rather different optical properties. The onset of the
absorption curve in the BOM with Si--O--Si angles of 144$^{\circ}$ and
158$^{\circ}$ do not differ that much from the one obtained from bulk-Si,
while the DBM and the BOM with 109$^{\circ}$, viz. the BOM with higher Si--O
bondlengths at the interface, give better optical properties. This is
additional indication that the interfacial atomic structure has to be
connected to the optical enhancement in the SLs, in agreement with
Kageshima's analysis.\cite{Kageshima} Likewise, it can be concluded from
Fig.\ \ref{tousabs} that bulk Si has poor optical properties compared to the
SLs, since the onset of absorption lags behind that for the two SLs and the
SDIG is much larger.

\section{Concluding Remarks}\label{ConcluSec}

We have used first-principles calculations to (i) study the electronic and
optical properties of Si/SiO$_2$ SL models and (ii) examine the applicability
of tight-binding calculations for these systems.

\begin{figure}[t]
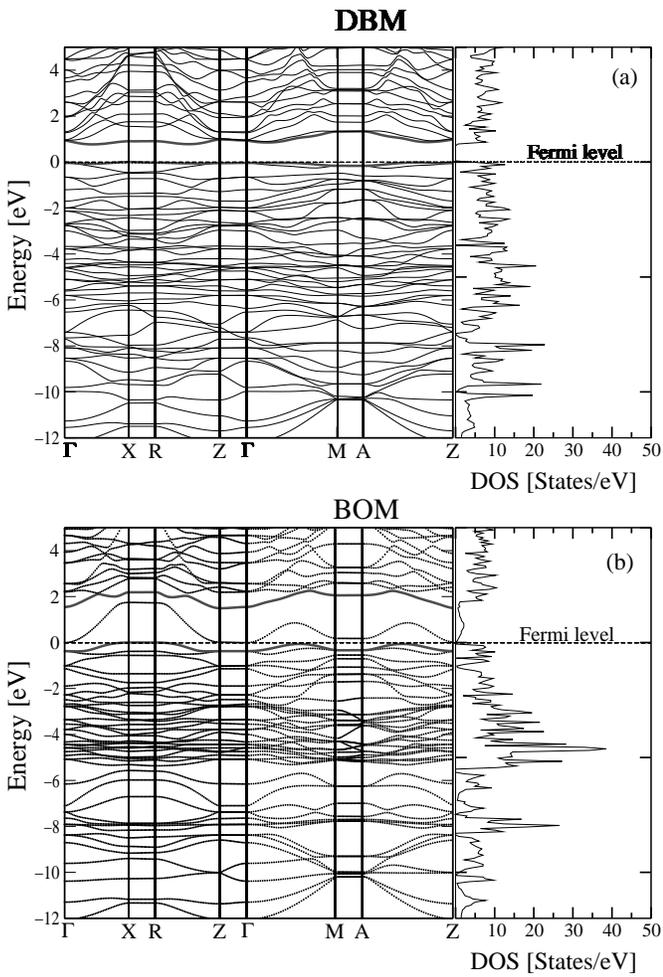

\includegraphics*[width=8.7cm]{Figs/bandDBM.eps}
\includegraphics*[width=8.7cm]{Figs/bandBOM144.eps}
\caption{
Band structure and DOS for (a) the DBM and (b) the BOM.
}
\label{DBMBOMband}
\end{figure}

\begin{figure}[t]
\includegraphics*[width=7cm]{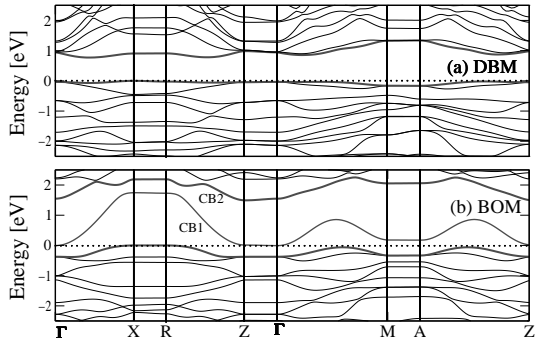}
\caption{
Detail of the band structure near the Fermi level for the DBM and the BOM.
}
\label{fermiband}
\end{figure}

Concerning the second point, our results indicate that, for both models, the
VBO value of 3.75 eV used in TB calculations is excessive. In addition, the
direct-transition nature of the gap from the TB calculations is incorrect.
The present calculations indicate that the band gap in both models is
indirect (albeit quasi-direct), in the red or infrared region of the
spectrum. Confinement is further confirmed by the essentially dispersionless
character of the electronic band structures in the growth direction (cf.
Figs~\ref{DBMBOMband} and \ref{fermiband}). Our calculations indicate, also,
that the SLs have enhanced optical properties as compared to pure Si.

The influence on the optical properties of the Si--SiO$_2$ interface has been
assessed from the absorption coefficient. This quantity depends critically on
the details of the interfacial structure. Realistic models are thus essential
for determining the electronic and optical properties. We are presently
studying improved models for this system.\cite{Tran}

\vspace{0.5cm}

{\it Acknowledgments} --
It is a pleasure to thank Gilles Abramovici, Ralf Meyer and Michel C\^ot\'e for
useful discussions. This work is supported by grants from the Natural
Sciences and Engineering Research Council (NSERC) of Canada and the ``Fonds
pour la formation de chercheurs et l'aide {\`a} la recherche'' (FCAR) of the
Province of Qu{\'e}bec. We are indebted to the ``R\'eseau qu\'eb\'ecois de calcul de
haute performance'' (RQCHP) for generous allocations of computer resources.

\newpage

\vspace{1cm}

\begin{table}
\caption
{
Possible (LDA-DFT) transition energies (in eV) for the two models (DBM and
BOM with Si--O--Si interfacial angle of 144$^{\circ}$). CB1, CB2, and VB refer
to the first and second conduction band, and the valence band, respectively.
For the DBM, $\Gamma X/2$ designates the conduction band minimum, which lies
approximately half-way between $\Gamma$ and $X$ points (cf.\ Fig.\
\protect\ref{fermiband}). The lowest direct and indirect transitions are
listed; the energy gaps, which are indirect, are indicated in boldface. For
the BOM, CB1 can be viewed as a gap state (see text), and the ``true''
LDA-gap is therefore 1.49 eV; for the DBM, this gap state is absent, i.e.,
CB1 is the true lowest band, and the LDA gap is 0.81 eV.
}
\begin{tabular}{l|lr|lr}
\hline
&   \multicolumn{2}{c|}{BOM} & \multicolumn{2}{c}{DBM}                        \\ \hline
                  & \mbox{} \hfill $X$ \hfill \mbox{}
                  & \mbox{} \hfill       $Z$\hfill \mbox{}
                  & \mbox{} \hfill  $\Gamma$\hfill \mbox{}
                  & \mbox{} \hfill $\Gamma X/2$\hfill \mbox{} \\ \hline
CB2               &  \hspace{3pt}2.1870  &      1.4932   &
    \mbox{} \hfill -----      & \mbox{} \hfill    -----       \\
CB1               &  \hspace{3pt}1.7447  &      0.0143   &
    0.9781   &    0.8050      \\
VB                &  \hspace{3pt}0.0000  &     -0.3754   &
    0.0000   &   -0.0369      \\ \hline
CB2$-$VB dir.     &  \hspace{3pt}2.1870  &      1.8686   &
    \mbox{} \hfill -----      & \mbox{} \hfill    -----      \\ \hline
CB2$-$VB ind.     &  \multicolumn{2}{c|}{CB2$_Z$--VB$_X$: {\bf 1.4932}} &
    \mbox{} \hfill  ----- &\mbox{} \hfill  ----- \\ \hline
CB1$-$VB dir.     &  \hspace{3pt}1.7447  &      0.3897   &
    0.9781   &    0.8419      \\
CB1$-$VB ind.     & \multicolumn{2}{c|}{CB1$_Z$--VB$_X$: {\bf 0.0143}} &
                    \multicolumn{2}{c}{CB1$_{\Gamma X/2}$--VB$_\Gamma$:
    {\bf 0.8050}} \\ \hline
%\hline
\end{tabular}
\label{gaps}
\end{table}

\end{document}